\documentclass{vgtc}      

\ifpdf                  
  \pdfoutput=1\relax                 
  \usepackage{graphicx}
  \DeclareGraphicsExtensions{.pdf,.png,.jpg,.jpeg}
\else
  \usepackage{graphicx}
  \DeclareGraphicsExtensions{.eps}
\fi

\graphicspath{{figures/}{pictures/}{images/}{./}} 

\usepackage{microtype}
\PassOptionsToPackage{warn}{textcomp}
\usepackage{textcomp}
\usepackage{mathptmx}
\usepackage{times}

\usepackage{cite}                      
\usepackage{tabu}               
\usepackage{booktabs}

\usepackage{paralist}
\usepackage{subfig}
\usepackage{color, soul}
\usepackage[font={scriptsize,sf},tableposition=top]{caption}
\usepackage{float}
\floatstyle{plaintop}
\restylefloat{table}
\usepackage{sparklines}
\usepackage{makecell}
\usepackage{multirow}
\usepackage{multicol}
\usepackage{colortbl}
\usepackage{pifont}
\usepackage{tabularx}
\usepackage{balance}
\usepackage{amsmath}
\usepackage[para,online,flushleft]{threeparttable}
\onlineid{1214}
\vgtccategory{Research}
\vgtcpapertype{application}
\vgtcinsertpkg

\newcommand{\Title}{
A Visual Analytics Approach to Debugging Cooperative,\\ Autonomous Multi-Robot Systems' Worldviews
}
\def\techname{DWC}
\def\mainview{Main View}
\def\worldview{Differential Worldview Comparison}
\title{\Title}

\author{
Suyun ``Sandra'' Bae\thanks{e-mail: suybae@ucdavis.edu}\\ \and 
Federico Rossi\thanks{e-mail: federico.rossi@jpl.nasa.gov}\\ 
\and 
Joshua Vander Hook\thanks{e-mail: hook@jpl.nasa.gov}\\
\and 
Scott Davidoff\thanks{e-mail: scott.davidoff@jpl.nasa.gov}\\ %
\and
Kwan-Liu Ma\thanks{e-mail: klma@ucdavis.edu}\\ %
}

\affiliation{ 
\scriptsize $\ast \P$ \text{ University of California, Davis}
\\
\scriptsize $\dagger \ddagger \S$ \text{ Jet Propulsion Laboratory, California Institute of Technology}
}

\shortauthortitle{Bae \MakeLowercase{\textit{et al.}}: \Title}

\abstract{
Autonomous multi-robot systems, where a team of robots shares information to perform tasks that are beyond an individual robot’s abilities, hold great promise for a number of applications, such as planetary exploration missions.
Each robot in a multi-robot system that uses the shared-world coordination paradigm autonomously schedules which robot should perform a given task, and when, using its \textit{worldview}--the robot’s internal representation of its belief about both its own state, and \textit{other} robots’ states.
A key problem for operators is that robots’ worldviews can fall out of sync (often due to weak communication links), leading to desynchronization of the robots’ scheduling decisions and inconsistent emergent behavior (e.g., tasks not performed, or performed by multiple robots).
Operators face the time-consuming and difficult task of making sense of the robots' scheduling decisions, detecting de-synchronizations, and pinpointing the cause by comparing every robot's worldview.
To address these challenges, we introduce MOSAIC Viewer, a visual analytics system that helps operators (i) make sense of the robots’ schedules and (ii) detect and conduct a root cause analysis of the robots' desynchronized worldviews. Over a year-long partnership with roboticists at the NASA Jet Propulsion Laboratory, we conduct a formative study to identify the necessary system design requirements and a qualitative evaluation with 12 roboticists. We find that MOSAIC Viewer is faster- and easier-to-use than the users’ current approaches, and it allows them to stitch low-level details to formulate a high-level understanding of the robots’ schedules and detect and pinpoint the cause of the desynchronized worldviews. 
}


\CCScatlist{ % not used in journal version
    \CCScat{I.3.8}{Computer Graphics}{Applications}%
}

\begin{document}
\firstsection{Introduction}
\maketitle

% for equations space
\setlength{\abovedisplayskip}{1pt}
\setlength{\belowdisplayskip}{1pt}

\firstsection{Introduction}

\maketitle
\begin{table*}[tb]\centering
\renewcommand{\arraystretch}{0.9}
\begin{tabular}{@{}l ll l@{}} \toprule 
 & \multicolumn{2}{c}{\textbf{Worldview Attribute Value}} \\ \cmidrule(r){2-3}
\textbf{Attribute} & Self & Others & \textbf{Data Structure}\\\midrule
Location & The robot's location & Presumed location of other robots & 2D Coordinates \\
Science Zone & The robot's classification of whether   & Presumed classification of whether  & Boolean Array \\
& it is in a science zone & other robots are in science zones & \\
Battery Level & The robot's battery & Presumed battery level of other robots & Ordinal Array \\
CPU Utilization & The robot's CPU level & Presumed CPU level of other robots 
 & Ordinal Array \\
Actions & The actions the robot is currently performing & Actions other robots are believed to be performing & Event Sequence \\ 
Communication & Bandwidth between self and other robots & Presumed bandwidths  between other pairs of robots & Graph\\ \bottomrule
\end{tabular}
\vspace{0.5em}
\caption{Attributes in \textit{every} agent's worldview. For every attribute, an agent has a value for itself and the presumed values for the other agents.}
\label{tab: worldview}
\vspace{-0.1cm}
\end{table*}

Autonomous multi-robot systems (MRS) are systems with two or more autonomous robots (often referred to as \textit{agents}), that coordinate and share information so as to perform tasks cooperatively. This cooperation, in particular, drives interest in their potential to perform highly complex tasks in diverse contexts from search and rescue in hazardous environments\cite{argrow2005uav,nagatani2011multirobot,gancet2010user} to team sports\cite{humphrey2006visualization, osawa1996robocup}, and even space missions\cite{rossi2018, seah2005multi}.
The complexity of these systems also introduces a problem of usability for operations---or operability. 
Researchers must monitor the behavior of individual agents as well as behavior that emerges from their cooperation\cite{holland2000emergence} and see how small changes to their systems affect the overall system performance.

One area where this emergent complexity can be particular challenging for operators is \emph{distributed scheduling} \cite{parker2007distributed}, i.e., cooperatively deciding which agent should perform a given task and when. To track even a single task on a single agent, operators need to understand and track inter-task dependencies the precedence constraints that the agents' schedule must satisfy (e.g., scientific data must be collected before it is analyzed and transmitted to Earth). With distributed scheduling, the effort required to understand the state of the system increases quadratically, as tasks assigned to one agent might be shared with or even performed by others. In addition, the overall system's behavior depends not only on each agent's individual state but also on each agent's belief about the state of other agents and of the environment (i.e., a \textit{worldview}\cite{halpern1990knowledge}).

Agents' worldviews introduce a second challenge that further complicates tracking agents' tasks, as worldviews can fall out of sync (often due to weak communication links). This leads to desynchronization of the agents' scheduling decisions and inconsistent emergent behavior (e.g., tasks not performed, or performed by multiple agents).
To debug these inconsistent behaviors, operators must pinpoint the source of the desynchronization by comparing every agent's worldview.
This process is not only critical for debugging and failure detection purposes, but also enormously time-consuming and difficult: operators need to examine the high-dimensional product of every attribute in every agent's worldview. 

While previous research has explored ways to represent the views of single agents using text\cite{de2007distributed, figueiredo2006multi, ndumu1999visualising} or superimposed over videos \cite{tanoto2006mpeg, annable2013nubugger}, we explore how a visual analytics approach\cite{cook2005illuminating} can encode the belief agents have of themselves and about the state of other agents.
To that end, we engaged in a year-long collaboration with a team of MRS researchers and operators at the NASA Jet Propulsion Laboratory (NASA JPL).
The collaboration began with a 10-week formative investigation to identify the core challenges of distributed scheduling, utilizing the MOSAIC distributed scheduling framework\cite{vander2019mars} as a laboratory to explore this objective. This work was followed by six months of iterative co-design with a core MOSAIC team member to produce MOSAIC Viewer, a visual analytics application that helps operators (i) make sense of the agents' schedules and (ii) detect and conduct a root cause analysis of the desynchronized worldviews. 
To compare worldviews, MOSAIC Viewer draws inspiration from the \textit{diff} algorithm, which is commonly used for text comparison\cite{hunt1976algorithm} to emphasize the differences of agents' worldview.
Lastly, we demonstrate the effectiveness of our
method and system with two case studies and evaluate the application through a qualitative study with 12 roboticists at JPL.
The study reveals MOSAIC Viewer is easier- and faster-to-use than the users' current text-based approaches. The study helps to explain how applications like MOSAIC Viewer can support worldview desynchronization debugging. In particular, from our evaluation, we find that our tool supports the following practices:

\begin{itemize}
    \itemsep0em 
    \item System speed and interactivity streamline higher-level analyses;
    \item Trust for summary displays grew with experience;
    \item Knowing how is not enough---users need to know ``why'' in order to back trace the root causes of the problem;
    \item Different sets of assumptions affect data interpretation.
\end{itemize}
\vspace{-0.3cm}
This particular design study \cite{sedlmair2012design} helps researchers understand the fit between the problem of MRS operators debugging desynchronized worldviews and MOSAIC Viewer.
In this paper we contribute: (i) a set of system design requirements based on a year-long formative study with domain experts in multi-robot systems; (ii) a visual analytics tool that helps operators understand and compare agents' worldviews with a comparison technique inspired from the \textit{diff} algorithm; and (iii) we characterize how the system supports effective troubleshooting, with evidence gathered from a study of the system. 
\section{Background}
To motivate and situate our work, we first discuss the specific challenges of supervising autonomous MRS and unpack the details of the MOSAIC distributed scheduling framework\cite{vander2019mars} that we use to explore desynchronized worldview debugging.

\subsection{Autonomous Multi-Robot Systems}
\label{sec:mrs}
In contrast to multi-agent systems \cite{michel2018multi} 
which are enacted entirely as software, in this work we focus on multi-robot systems that have to negotiate with real-world constraints (e.g., limited and time-varying communication bandwidth and dynamic battery levels) that are often not considered in the multi-agent systems literature \cite{kardas2012design, luo2002multi, burmeister1997application}.

While some MRS researchers investigate MRS that utilize explicit \cite{simmons2000coordination} or centralized \cite{yamashita2003motion} coordination, we focus on MRS that use \emph{shared-world}  coordination\cite{arai2002advances,entin1999adaptive}, which have proven to be highly popular in field applications \cite{baxter2007multi, rossi2018routing} due to its simplicity, scalability, and resiliency. 
In this approach, every robotic agent has a \textit{worldview}\cite{halpern1990knowledge}---an internal representation of the world and of the other agents’ states---that is updated through constant communication with other agents.
In our case, \autoref{tab: worldview} summarizes the different attributes found in an agent's worldview within the MOSAIC framework (described in \autoref{sec:mosaic}).
Based on its own worldview, each agent independently computes the optimal strategy for \emph{all} agents, and then executes its own part of the computed strategy.
If all agents have the same worldview, this results in coordinated behavior.

While a shared-world approach introduces many benefits, a key complexity it introduces is that, if the agents are unable to communicate with each other, their worldviews can fall out of sync, resulting in uncoordinated decisions (e.g., a task may be performed by two agents, or it may not be performed at all).
This issue is especially problematic in harsh environments such as underground caves\cite{kunsei2018improved, emslie1975theory}, where ensuring constant reliable communication is infeasible. Therefore, in order to understand the overall behavior of MRS, it is critical to understand the worldview of each agent.
Furthermore, in order to mitigate the effect of worldview desynchronization, an operator must be able to identify the cause (e.g., slow propagation of information on low-bandwidth data links or the failure of an agent's radio) to plan for corrective action.
The concurrent, distributed, and complex components of MRS makes the debugging process significantly difficult, and previous research has identified that these tasks require considerable attention\cite{garcia1984debugging} and would benefit from more appropriate, effective tools \cite{van1995debugging}. 

\subsection{MOSAIC Distributed Scheduling}
\label{sec:mosaic}
Within the MOSAIC distributed scheduling framework, each agent can perform a set of \textbf{navigation} tasks and \textbf{science} tasks.
Navigation tasks, which model activities such as localization and path planning, are \textit{mandatory}, and all agents must perform them.
Science tasks, which model collection and analysis of scientific observables, are \textit{optional}.
Though individual science tasks are not required for mission success, the objective of the MOSAIC framework is to perform as many science tasks as possible.
However, an agent can perform science tasks only if it  has the time and energy resources to also guarantee the execution of the mandatory navigation tasks.

Both navigation tasks and science tasks have precedence constraints, enforcing that tasks must be accomplished in sequence.
With science tasks, performing analysis of scientific measurements requires that the data be collected first. With navigation tasks, performing localization through visual odometry requires collecting camera images first. 
Hence, we will refer to navigation tasks and science tasks as a ``chain of tasks'' (i.e., several tasks with a chain of inter-task dependencies).
A key advantage of MRS is that agents do not need to accomplish each task all by themselves---they may receive assistance from other agents. 
An agent in a science zone may request assistance with its navigation tasks in order to free up computational resources for science tasks.
Certain computational tasks, such as performing visual odometry and analyzing data, are relocatable to other agents.
However, not all tasks are relocatable---for instance, tasks requiring the use of an agent's hardware resources (e.g., capturing images or collecting scientific measurements) are not.
The MOSAIC scheduler takes all these constraints into account and computes (i) what optional tasks can be performed and (ii) which agents should perform relocatable tasks based on the agents' capabilities and communication links between the agents \cite{vander2019mars}.

\begin{table*}[htb]
  \renewcommand{\arraystretch}{0.8}
  \small
  \centering
  \begin{threeparttable}
  \begin{tabular}{l
                  >{\raggedright}
                  p{0.4in}
                  >{\raggedright}
                  p{0.6in}
                  >{\raggedright}
                  p{1.2in}
                  ccccccccccc}

    \toprule

    \multicolumn{2}{l}{\multirow{3}{*}{\textbf{Participants}} } &
    \multirow{3}{*}{\makecell[l]{\textbf{MOSAIC} \\\textbf{Affiliation}}} &
    \multirow{3}{*}{\makecell[l]{\textbf{Years of Full-Time} \\\textbf{Professional Experience}}} &
    \multicolumn{5}{c}{\textbf{Current Tools}} & &
    \multicolumn{4}{c}{\textbf{Participation}}\\
    % & &\multirow{3}{*}{\textbf{Evaluation}}\\
    \arrayrulecolor{black!30}\cmidrule(l){5-9}
    \cmidrule(l){11-14}
     & & & & \textbf{PLT}\tnote{1}&\textbf{RT}\tnote{2}& \textbf{CLI}\tnote{3} & \textbf{DBT}\tnote{4}&\textbf{DB-LT}\tnote{5}& &\textbf{FS\tnote{6}} & \textbf{Co-Design} &
     \textbf{FE\tnote{7}}&
     \textbf{User Study}\\
    \arrayrulecolor{black}\midrule
    
     ~ &
    P0 &
    Core &
    1 -- 5 years &
    \ding{51} & \ding{51} & \ding{51} & -- & -- &&
    \ding{51} & \ding{51} & \ding{51} & --\\

    ~ & 
    P1 &
    Core &
    1 -- 5 years &
    -- & \ding{51} & -- & -- & -- &&
    \ding{51} & -- & -- & \ding{51}\\

    ~ &
    P2 &
    Core &
    Less than 1 year &
    -- & -- & -- & -- & \ding{51} &&
    -- & -- & -- & \ding{51}\\

% %P2:PostgreSQL, 20hrs/week, \textcolor{red}{To support sharing information among agents}

    ~ &
    P3 &
    Non-Core &
    5 -- 10 years &
    \ding{51} & -- & -- & -- & -- &&
    \ding{51} & -- & -- &\ding{51}\\
    
% %P3:Matlab, 20hrs/week, To support guidance, navigation, control, and estimation using multi-agent systems and swarms &

    ~ &
    P4 &
    Non-Core &
    Less than 1 year &
    \ding{51} & -- & \ding{51} & -- & -- &&
    -- & -- & -- & \ding{51}\\

% % P4: Log extractions and simulation tools, 5-10 hrs/week, To support debugging new multi-agent system software 

    ~ &
    P5 &
    Core &
    10 - 15 years &
    -- & -- & \ding{51} & -- & -- &&
    \ding{51} & -- & \ding{51} &\ding{51}\\
    
% % P5: Terminal logs, 2hrs/week, To support simulation 

    ~ &
    P6 &
    Non-Core &
    1 -- 5 years &
    \ding{51} & \ding{51} & -- & -- & -- &&
    \ding{51} & -- & -- &\ding{51}\\

% %  P6 MATLAB Plots and RViz/Gazeo on ROS, 4hrs/week, To support validation tasks 

    ~ &
    P7 &
    Non-Core &
    1 -- 5 years &
    -- & \ding{51} & \ding{51} & \ding{51} & -- &&
    -- & -- & -- &\ding{51}\\

% % P7: (1) RViz, 2hrs/ week, simulation or logged data, (2) Vim, 4hrs/ week, Reading debugging logs, (3) gdb, 1hr/week, Backtraces from Unrecoverable programm errors

    ~ &
    P8 &
    Core &
    1 -- 5 years &
    \ding{51} & -- & \ding{51} & -- & -- &&
    \ding{51} & -- & -- &\ding{51}\\
    
% % P8: Graphs and terminal logs, 20 hrs/week, debugging

    ~ &
    P9 &
    Non-Core &
    1 -- 5 years &
    -- & \ding{51} & -- & -- & -- &&
    -- & -- & -- &\ding{51}\\

% % P9: ROS and RViz, 2 -- 5 hrs/week, Autonomous Puffer Task 

    ~ &
    P10 &
    Non-Core &
    More than 15 years &
    -- & \ding{51} & \ding{51} & -- & -- &&
    -- & -- & -- &\ding{51}\\
    
% % P10:(1) Terminal logs, ??,  debug / find issues; (2) CTB RVIZ, ??, to visualize and understand what and where each agent is doing (3) MASSIVE, ??, position on the map, tasks allocated, collision avoidance data.

    ~ &
    P11 &
    Non-Core &
    1 -- 5 years &
     -- & \ding{51} & -- & -- & -- &&
    -- & -- & -- &\ding{51}\\

% % P11:RViz, 10hrs/week, To support debugging and Monitoring

    ~ &
    P12 &
    Core &
    10 -- 15 years &
    -- & -- & \ding{51} & -- & -- &&
    \ding{51} & -- & \ding{51} & \ding{51}\\
    
%P12: CLI, 5hrs/week, \textcolor{red}{5 tasks}

    \arrayrulecolor{black}\bottomrule

  \end{tabular}
  \begin{tablenotes}
    \linespread{0.5}\small   
    \item[1] MATLAB/Matplotlib (ad-hoc scriping languages and plotting tools);
    \item[2] RVIZ and ROS-based plotting tools\cite{quigley2009ros};
    \item[3] Command-line (CLI) logging tools; 
    \item[4] Command-line debugging and backtracing tools (GDB); 
    \item[5] Database-backed logging tools;
    \item[6] Formative Study;
    \item[7] Formative Evaluation
    \end{tablenotes}
    \end{threeparttable}
  \caption{Summary of the participants' background, current tools and the extent they participated in the year-long formative study.}
  \label{tab:users}
\end{table*}

In this paper, we consider datasets with ten agents generated by running the MOSAIC scheduler in the loop with a multi-robot simulator that captures the availability of science zones, robot battery levels, and communication link bandwidths, which reflects the standard practice in robotics research\cite{vander2019mars, vander2019autonomous, pinciroli2012argos}.
This level of simulation fidelity is well-matched with the level of abstraction at which MOSAIC operates; while field testing may present different underlying causes for worldview desynchronization, the effect on the data used in this paper (i.e., disagreement between the agents' world views) would be indistinguishable from the output of the simulations. Accordingly, the use of a simulator has a negligible impact on the fidelity of worldview debugging.
In these datasets, each agent has six attributes in its worldview (\autoref{tab: worldview}) and is endowed with an agent ID. Each agent wishes to perform  three mandatory navigation tasks; agents in ``science zones'' also wish to perform three optional science tasks. Each set of tasks has a chain of dependency constraints. 
One agent---the base station---is a special agent that does not need to perform the navigation or science tasks and is equipped with a faster processor. Its purpose is to help the other agents with its computing capabilities.
Lastly, we remark that the scale of ten agents is representative of proposed extra-planetary (i.e., outside of Earth) MRS mission concepts under consideration for the next decade \cite{board2012vision,rossi2018, kasper2019sun, karras2017puffer}.
\section{Related Work}
We present MRS supervision tools used in the robotics community and promising visualization techniques for worldview debugging.

\subsection{Scheduling and Timeline Views in Robotics}
In addition to the various multi-purpose robotics toolkits\cite{quigley2009ros}, specialized visual analytics tools have been developed to track robots' task completion\cite{ndumu1999visualising, szekely2001interfaces, taylor2002vista}. These tools include timeline views that are often organized around two underlying data types. First, we observe ``agent-centric'' (AC) timelines\cite{jin2008vizscript, cummings2005managing}, which map timeline rows to individual agents, foregrounding the tasks performed by each agent (either for themselves or to assist other agents). In contrast, the second approach, ``task-centric'' (TC) \cite{ruff2013human}, organizes timeline rows around tasks and their dependencies and focuses on when tasks are completed, rather than \textit{who} is executing them.

We find AC timelines to be an incomplete solution for the problem of distributed scheduling, where tasks offloaded to other agents can be difficult to trace. We find a similar limitation with TC timelines for distributed scheduling problems, where backtracing tasks with dependencies can be difficult to explain failures. To that end, MOSAIC Viewer uses an AC-TC hybrid timeline. The timeline includes glyphs that visually encapsulate the completion status of individual tasks, and foregrounds task dependencies using interactions\cite{wongsuphasawat2011lifeflow, jo2014livegantt, fails2006visual} (see \autoref{sec:interactions} for more details about the timeline). Even with these adaptations, the timeline view is necessary but not sufficient to support the many tasks robotics researchers and operators face when debugging unexpected behaviors due to worldview desynchronization. To complete the application, we turn our attention to views designed to support worldview debugging.

\subsection{Worldview Debugging}
To mitigate the effect of worldview desynchronization, an operator must be able to identify the root cause of the desynchronization condition (\autoref{sec:mrs}). Researchers have investigated methods that display worldview state variables using structured text, through logfile analysis\cite{figueiredo2006multi} and watchpoints\cite{de2007distributed}, similar to those found in software development IDE's.
These initial explorations provide utility, but at the same time require considerable attention and focus \cite{tullis1984predicting, preece1994human}. One objective of our research is to explore a more expressive and lower cognitive load approach for users to examine the high-dimensional product of agents' worldviews. Visual tracking\cite{annable2013nubugger, tanoto2006mpeg}, which overlays line graphs\cite{annable2013nubugger} and glyphs\cite{tanoto2006mpeg} that describe the agent's state on top of a video of agents performing tasks, is another approach. While this approach can effectively show the state of individual agents, MRS worldview debugging requires operators to understand agents' beliefs about \textit{other} agents' states as well.

To the best of our knowledge, we know of no research within the visualization community that has explicitly looked at the problem domain of distributed MRS worldview debugging. However, researchers have explored various representations and interaction strategies for data with similar underlying representations.
For example, as worldviews have multiple attributes, we build on visual comparison techniques rooted in multivariate data research \cite{liu2016visualizing}. In particular, we rely on visual comparison techniques\cite{gleicher2011visual} that include explicit encoding to represent the system's consensus of an agent's view of its own state, and juxtaposition to highlight the differences among worldviews.
The same strategies to compare parameters of a dataset have been applied across a number of domains, from malware sampling\cite{gove2014seem} to time series data\cite{niederer2017taco}.

Another thread that defines analytics tools that perform multivariate comparisons is the actual algorithm they select to highlight parameter differences. DiffMatrix\cite{song2012diffmatrix}, for example, highlights the difference between two parameter values using the arithmetic subtraction operators, while OnSet\cite{sadana2014onset} uses the union and intersection set operators. Our work, like Vdiff\cite{barnes1988developing} and TACO\cite{niederer2017taco}, applies the diff algorithm\cite{hunt1976algorithm} from text processing (see \autoref{sec:application:dwc} for details).
The system we present, overall, extends the work on multivariate comparison into the domain of distributed MRS worldview debugging and contributes a detailed analysis of the fit of our approach to that domain.
We now turn our attention to the formative research that informed our understanding of the problem.

\section{Formative Study}
MOSAIC Viewer is the product of a year-long engagement, organized around three distinct phases, with its users. \autoref{tab:users} summarizes the phases of the engagement and describes how each of the 13 participants engaged with this project. (Note: P0 is a superuser who guided the design process, but did not participate in the user study).

This section describes the first and second phase: a 10-week formative study with 7 domain experts and a 6-month co-design study with 1 domain expert. The third phase (\autoref{sec:userstudy}) is a formal user study that evaluates how well MOSAIC Viewer supports visual debugging for worldview desynchronization.

\subsection{Approach}
The objectives of the first phase, a formative study, was to gain a deeper understanding of the core challenges of distributed scheduling and the users' needs. The study was organized around an initial contextual inquiry\cite{holtzblatt1997contextual}, which allowed the research team to observe the roboticists at work in their own environments.
For four weeks, the protocol consisted of six sessions of 90 minutes of semi-structured interviews and an artifact walk-through where roboticists shared the tools and processes that define their MRS work practice.
The research team took notes, and captured images and video recordings to highlight observations for post-analysis.

With a preliminary understanding of the problem domain, the research team then elaborated 12 paper prototypes\cite{rettig1994prototyping} in order to test initial ideas early and quickly. After four rounds of user testing with all users, these prototypes evolved to higher-fidelity code-based prototypes with real MRS data. 
From these prototyping sessions, we identified that comparing agents' worldviews is a core component we needed to address in order to help roboticists make sense of the agents' schedules. Then in the second phase, we engaged in 6 months of detailed co-design sessions\cite{halloran2006unfolding} with P0 to unpack the specific challenges of worldview comparisons.
We found that no existing visualizations met the needs of the problem, and we collaborated with P0 to iteratively prototype different visual representations that compare agents' worldviews.
The past designs can be found in the supplementary material.
After identifying a potential design, we conducted a formative evaluation\cite{elmqvist2015patterns} with three roboticists, where we revised the system design based on their feedback, such as introducing the hierarchical interactions that we list in \autoref{sec:interactions}.

\begin{figure}[tb]
	\centering
	\captionsetup{farskip=0pt}% <--- no gap at the 
	\includegraphics[width=\columnwidth]{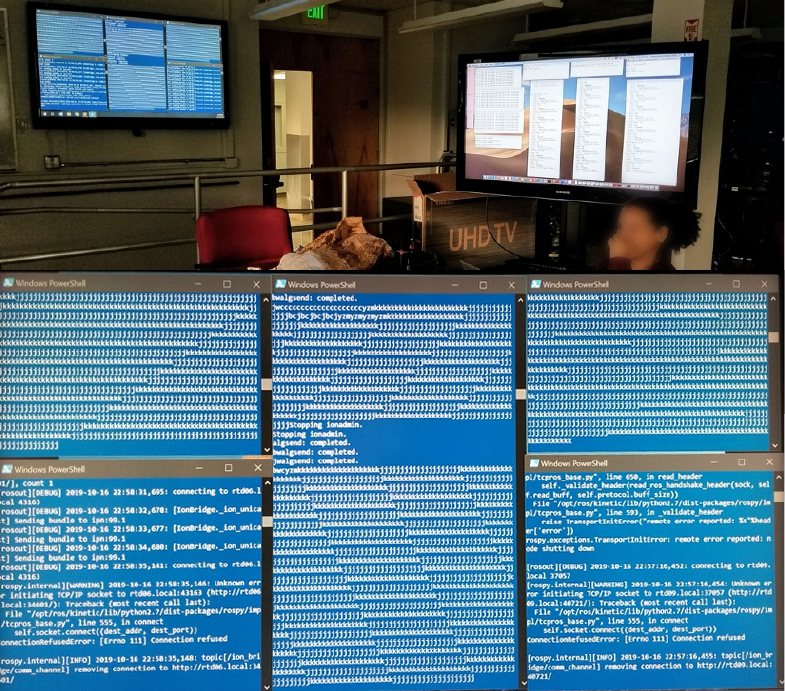}
    \caption{A 50-inch and a 70-inch display. The 15 terminal windows allows P1 and 2 other researchers (not pictured) to debug 3 robots.}
    \vspace{-0.2cm}
	\label{fig:monitors}
\end{figure}
\subsection{Findings}
The materials captured during phase 1 were annotated with themes, which were then grouped using an emergent theme analysis\cite{altheide2008emergent}.

One key analysis takeaway reveals roboticists invested substantial amounts of effort and time to track the state of each agent's worldview in order to interpret which tasks are scheduled for which robot.
To illustrate, we observed examples where roboticists would need to open 3 or 4 terminal windows of messaging logs per agent and then would recruit multiple expert colleagues who would each be tasked to monitor a single agent on their own screens (\autoref{fig:monitors}).
To elaborate a shared understanding of all the worldviews, including the divergences, the roboticists would collectively, as P1 shares, ``shout out the state of what they’re seeing'' as they visually scan their respective displays to find relevant information.

Roboticists must first assess worldview synchronization as this dictates the flow for the rest of their analysis.
If roboticists conclude that the worldviews are synchronized, they would then assess the system's performance (i.e., how many tasks are being performed, and by whom).
To perform this objective, we observed researchers first engaged in a TC perspective to quickly determine the state of the agent’s task: they did not seek the low-level details of how a task has been accomplished.
Conversely, if an agent failed, researchers switched to an AC perspective to seek for the low-level details of an agent's activities that explained the failure.
However, if roboticists concluded that agents' worldviews are desynchronized, they would transition into a debugging mode to plan for corrective action (i.e., determine who is out of sync and reason why the desynchronization occurred).
Roboticists would read out single states within each of their robots' worldviews again and check for agreement about each state before moving onto the next worldview.

\subsection{Design Requirements}
Phase 1 identified that worldview debugging requires considerable cognitive effort, supporting our decision to investigate how a visual analytics tool can support these  sets of tasks. In phase 2, over months of iterative co-design with P0, we elaborated on the specific challenges researchers face when debugging MRS systems (summarized in \autoref{tab:goals}). We use these questions to devise a set of requirements on how a visual analytics system can provide the needed support:\\
\begin{table}[tb]
\resizebox{\columnwidth}{!}{%
\begin{tabular}{p{3cm}@{\hskip 0.1in}p{6.5cm}}\toprule
        \multicolumn{1}{c}{Goal}&  \multicolumn{1}{c}{Subgoal}\\\midrule 
        \multirow{2}{*}{\makecell[l]{Assess worldview\\ synchronization}} & 1. If desynchronized, who is out of sync with whom? \\
        & 2. What is the cause of the desynchronization? \\
        \addlinespace[1em]
        \multirow{3}{*}[-0.5em]{\makecell[l]{Make sense of the \\scheduler's output}} & 1. Who is doing what? \\
        & 2. Are agents accomplishing their navigation tasks?\\
        & 3. Are agents maximizing their science tasks?\\
        \bottomrule
\end{tabular}
}
\caption{Researcher's goals and subgoals.}
\label{tab:goals}
\end{table}

\begin{compactdesc}
  \item[\textbf{R1.}] \textbf{Display worldview synchronization state:}
    Determining if the worldviews are in sync is the first step to an in-depth analysis.
    The system's visual encoding should indicate whether agents are in sync and highlight those who are not.
  \item[\textbf{R2.}] \textbf{Support system performance assessment:}
  System performance is based on the number of accomplished tasks. With $n$ agents, the number of tasks and task transferring can be high. To track who is doing what, the system should (i) differentiate the different tasks agents can perform and (ii) offer the flexibility to switch analytical perspectives (i.e., AC, TC).
  \item[\textbf{R3.}] \textbf{Show the differences and similarities of the worldviews:}
    When worldview desynchronization occurs (R1), researchers will need to detect who is out of sync in order to plan for corrective action. Detecting out of sync agents requires understanding the similarities and differences of all the worldviews.
  \item[\textbf{R4.}] \textbf{Help conduct a root cause analysis of the desynchronized worldviews:}
  After determining which agents are out of sync (R3), operators will need to conduct a root cause analysis to reason on why this desynchronization occurred and plan for corrective action. The system should provide the ability for users to collate and fuse pieces of information in order to properly diagnose the desynchronization condition.
\end{compactdesc}
\begin{figure*}[t]
    \centering
    \includegraphics[width=\textwidth]{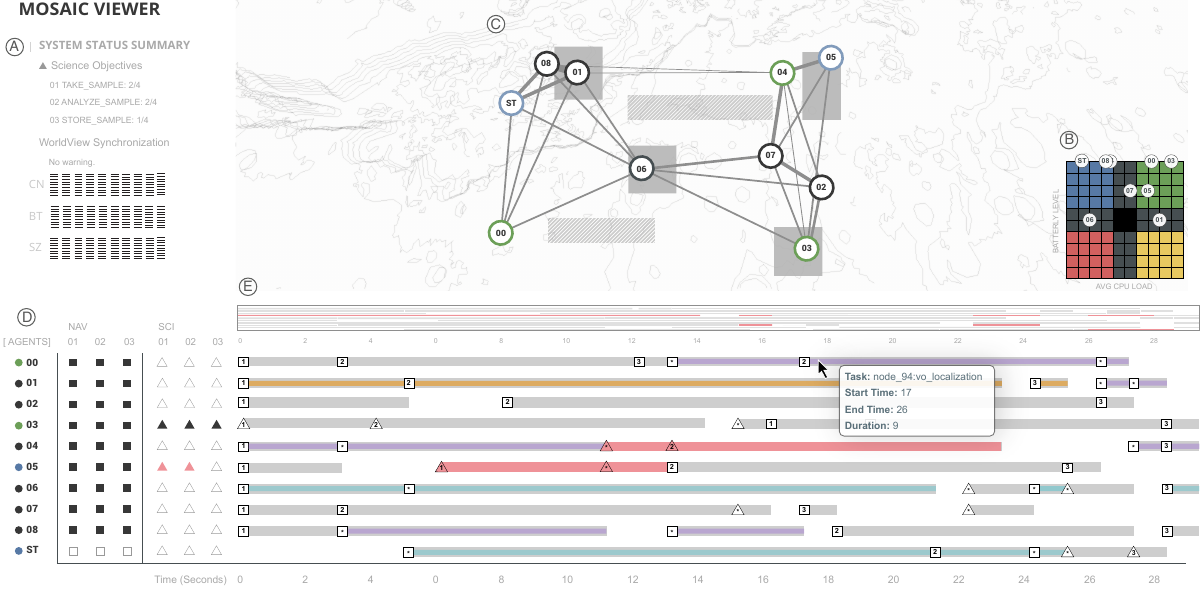}
    \caption{The operator is using the Main View to evaluate the system's performance.
    (a) Summary overview shows state of the science objective and the worldview synchronization
    (b) Scatterplot abstracts the ``behavior'' of agents with the x- and y-axis encoding average CPU load and battery level, respectively. 
    (c) Graph depicts the agents' location and communication links
    (d) Task Abstraction provides a task-centric perspective of each agent's task. Squares represent navigation tasks. Triangles represent science tasks
    (e) Timeline shows agents' activities. Agent 5's science chain of task is highlighted in pink. (Note: other agents' timelines are highlighted in different colors for illustrative purposes for the case study in \autoref{sec:case})}
    \label{fig:main}
\end{figure*}

\section{Application}
\begin{figure*}[t]
    \centering
    \includegraphics[width=\textwidth]{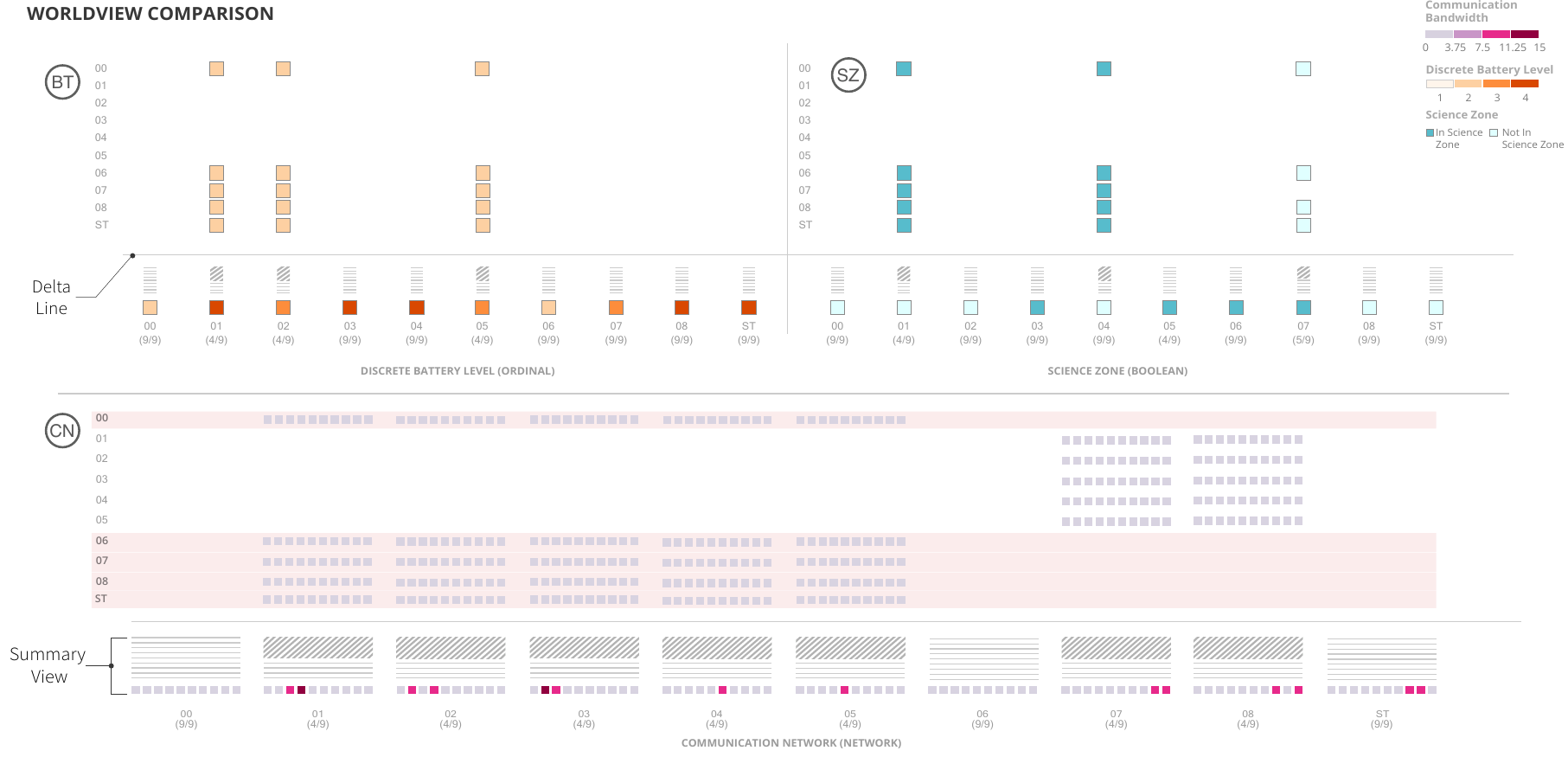}
    \caption{\worldview: (BT) Battery Level Panel, (SZ) Science Zone Panel, and (CN) Communication Network Panel.} 
    \label{fig:wvc}
    \vspace{-0.5em}
\end{figure*}

To support analyzing and debugging MRS, we design a system composed of two components: \mainview\ (\autoref{fig:main}) and \worldview\ (DWC) (\autoref{fig:wvc}).
The two components are laid side-by-side, and users can first use the \mainview\ to assess whether agents' worldviews are synchronized. If they conclude worldviews are synchronized, they can proceed to assess the system performance with the various views displayed in the \mainview. Otherwise, users can use DWC to identify worldview differences. After identifying and selecting misbehaving agents in DWC, users can engage with the \mainview\ to conduct a root cause analysis.

\subsection{Main View}
The \mainview\ (\autoref{fig:main}) consists of a summary overview, scatterplot, graph, task abstraction, and a timeline. 

\textbf{Summary Overview.} The summary overview (\autoref{fig:main}a) provides a quick assessment of agents' performance of science tasks and worldview synchronization \textbf{(R1, R2)}.
As navigation tasks are required, the operator can infer how well the MOSAIC scheduler is performing by assessing the number of optional science tasks that agents can accomplish.
Thus, for each task in the chain of science tasks, we display how many agents have executed the respective task as a fraction of all eligible agents. 
In \autoref{fig:main}a, 2 out of 4 agents have accomplished the first two science tasks. Of the two, only one has accomplished the third science task (either on board or by delegating it).
Below, \textit{Worldview Synchronization} provides an overview of the agents' worldview synchronization. To help users determine worldview synchronization, if the system detects a desynchronization, it outputs a warning (\textbf{R1}). Users can then turn to \techname\ for further analysis.
To show \textit{what} is desynchronized, the Worldview Synchronization displays \techname's visual representation for three worldview attributes (`CN': Communication Network, `BT': Battery Level, and `SZ': Science Zone). We discuss this further in \autoref{sec:application:dwc}.

\textbf{Scatterplot.} 
The scatterplot (\autoref{fig:main}b) is broken into four quadrants and the x- and y-axis shows average CPU load and battery level, respectively. 
The quadrants are colored with four different categorical colors to help operators abstract agents' behavior at-a-glance with respect to the current time step \textbf{(R2).}
CPU load, in particular, shows how busy a given agent is. 
This information in addition to battery level provides two indications: (1) it identifies over-subscribed agents and ``bottlenecks''; (2) it identifies agents that are doing the majority of the work for the overall system.
For example, agents in the blue quadrant are considered ``lazy''. 
With a high battery level and low average CPU load, this suggests that  these agents could be given more tasks if the communication topology allowed for it. 
In contrast, agents in the yellow quadrant have a low battery level and high average CPU load--indicating they are ``overworked''.
The middle portion of the scatterplot is colored grey, with the center colored black, in order to emphasize extreme behavior. 

\textbf{Graph.} 
In the graph (\autoref{fig:main}c), each agent is represented as a circle with their agent number in the center.
The base station is denoted as `ST'.
The colored ring correlates to which quadrant an agent is positioned in the scatterplot. 
The edges between agents represent its communication links, and 
the edge weight encodes the available communication bandwidth.
The graph has two regions to represent the agent's physical environment. 
Solid dark grey filled-in regions are the `science zones', while the diagonal hashed regions represent `communication cut-off zones'.
Communication links that cross a communication cut-off zone are severed, simulating the effect of line-of-sight of obstructions (e.g., hills). 
The graph provides basic interactions to zoom-in, pan, and see details on-demand (i.e., tooltip) when hovering over the nodes or edges.
Users can also click on an agent and see its immediate edges in pink, while the opacity of the other edges in the network is lowered.

\begin{figure*}[t]
    \centering
    \includegraphics[width=\textwidth]{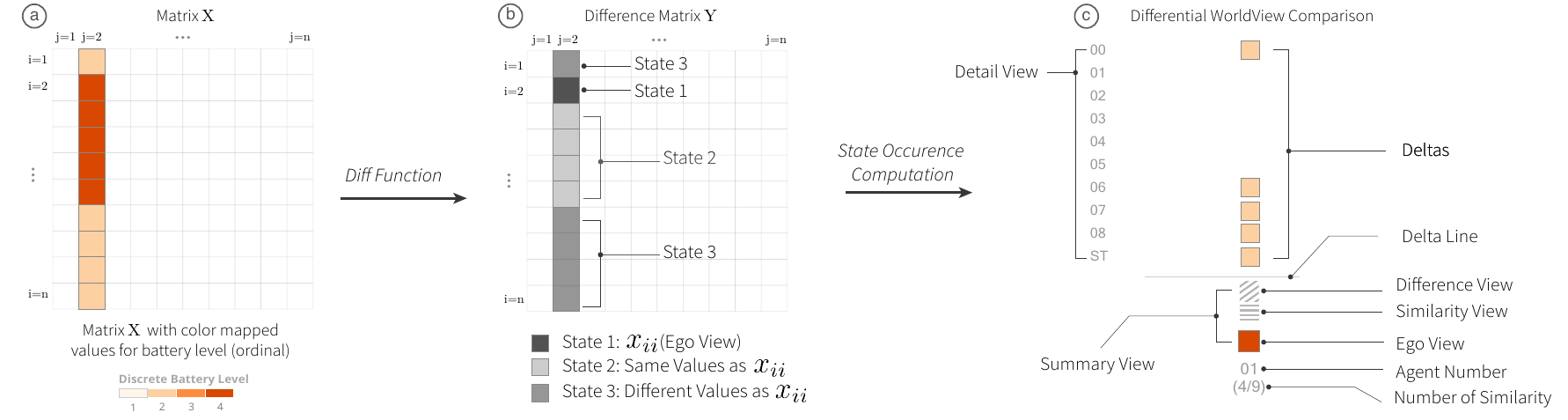}
    \caption{The process from the raw data to \techname. This figure focuses on Agent 2's battery level (\autoref{fig:wvc}). (a) The agents' view of a single attribute is represented as \textbf{X}, a $n{\times}n$ matrix.
    After applying our variant of the diff function to \textbf{X}, we obtain the Difference Matrix \textbf{Y} (b) where each entry is categorized as either State 1, 2, or 3. We compute the number of occurrences for each state. (c) The information is transformed into  \techname.}
    \label{fig:visual-encoding}
    \vspace{-1em}
\end{figure*}

\textbf{Task Abstraction and Timeline.} 
In  the Task Abstraction (\autoref{fig:main}d), each row represents an agent, and a colored circle next to the agent name corresponds to the quadrant the agent is respectively placed in the scatterplot.
To help distinguish tasks, there are two sets of shapes, where squares represent navigation tasks and triangles represent science tasks \textbf{(R2)}. 
Each chain of tasks is composed of three steps, and for every agent, the respective shape is filled in if the task has been accomplished.
This allows operators to quickly assess if agents have accomplished their navigation tasks, and how many science tasks have been performed from a TC perspective \textbf{(R2)}.
However, this task abstraction does not tell \textit{which} agent has completed the task or \textit{when} it has been accomplished.
To that end, operators can use the timeline, which shows each individual agent's activity (\autoref{fig:main}e) as horizontal bars in seconds.
The length of a bar represents the duration of a task and the positioning maps when a task has started and ended.
The shape at the beginning of each task represents the task type (navigation or science), and inside each shape is either a number that represents the nth-step of its respective chain of tasks or an asterisk symbol (*) to denote task relocation.
At first sight, it is not evidently clear if a task in an agent's timeline may have been relocated from another agent for assistance (e.g., Agent 0 is performing Agent 4's second navigation task at $t=17$).
To help provide a closer inspection, users can brush the range on the timeline with a mini-timeline or see more details of a task with a tooltip.
The tooltip displays the name, the start and end time of the task, and who the task belongs to. 
More interactions are listed in \autoref{sec:interactions}.

\subsection{Differential WorldView Comparison}
\label{sec:application:dwc}
%P: (1) Provide an overview of the Diff WorldView and how we are basing it off of the diff algorithm and (2) Explain why we're comparing based on the ego view.
Differential Worldview Comparison (DWC) enacts the concept of the \textit{diff} algorithm for text comparison\cite{hunt1976algorithm} to compare the agents' worldviews.
For each line in two text files, the diff algorithm either generates nothing, where the two lines are the same, or shows a side-by-side comparison of the two lines, where they differ.
Analogously, for each attribute in the agents' worldviews, we introduce a variant of the diff function that compares each agent's presumed value for an agent's attribute (e.g., agent B's presumed value for agent A's location) with the ego value of the said attribute (i.e., the value the respective agent has determined for itself. See \autoref{tab: worldview}).
We compare to an agent's ego value because we follow the strong assumption that every agent knows its state the best. Hence, the agents' ego value for each attribute acts as a form of comparison.
If an agent's presumed value corresponds to the respective ego value, \techname\ shows nothing; however, if they differ, \techname\ shows the presumed value.
\autoref{fig:visual-encoding} provides a walk-through of the process from the raw data to \techname.

Of the six attributes listed in \autoref{tab: worldview}, we focus on the three attributes (battery level, presence in a science zone, and communication network bandwidths) that have a \textit{direct} impact on the agents' scheduling decisions and are able to explain the majority of the desynchronization scenarios. 
For each agent $i$, each individual worldview attribute (e.g., battery level) can be represented as a 1D-array $x_i$ with $n$ entries, where $n$ represents the total number of agents (\autoref{fig:visual-encoding}a). 
The $j$-th entry in the array, $x_{ij}$ represents agent $i$'s belief about the state of agent $j$'s attribute. For battery level, the entry is an integer; for science zone, it is a boolean; and for communications, the entry is a list of bandwidths from agent $j$ to all other agents.

Once all $n$ arrays are concatenated, this becomes an $n{\times}n$ matrix \textbf{X}.
In \textbf{X}, row $i$ represents agent $i$'s beliefs about the attribute, and the entry $x_{ij}$ represents agent $i$'s belief about the state of agent $j$'s attribute value.
The value $x_{ii}$ is denoted as as the \emph{ego value} for the attribute, as this entry represents what agent $i$ thinks about itself.
For example, the highlighted column in \autoref{fig:visual-encoding}a represents every agent's belief about the state of Agent 2's battery level (encoded by color), and $x_{22}$ represents Agent 2's ego value for its battery level.

% \vspace{-1em}
After compiling \textbf{X} for each attribute, we apply our variant of the \textit{diff} function to every column $j$ in \textbf{X}, comparing every entry ($x_{ij}$) to the ego value:
\begin{equation}
     \mathrm{diff}(x_{ii}, x_{ij}) =
 \begin{cases}
      \text{None}, & \text{if}\ x_{ii} = x_{ij}\\
      x_{ij}, & \text{otherwise}
  \end{cases}
  \label{eq:1}
\end{equation}
If $x_{ii} = x_{ij}$, the function does not return anything; otherwise the value is $x_{ij}$ (i.e., agent $i$'s presumed value for agent $j$).
Once we compute the diff function for every column, the output values are used to create the Difference Matrix \textbf{Y} (\autoref{fig:visual-encoding}b). 
In \textbf{Y}, each entry is categorized as one of three states: State 1, State 2, or State 3.
State 1 is the ego view ($y_{ii}$), (ii) State 2 are entries ($y_{ij} = \text{None}$) that agree with the ego view, and (iii) State 3 are those ($y_{ij} \neq \text{None}$) that disagree with the ego view.
Then, for every column $j$ in \textbf{Y}, we compute the Similarity and Difference Sum to count the number of entries  labeled as State 2 and State 3, respectively:
\begin{equation}
    \mathrm{Similarity\ Sum} =\sum_{i=1}^{n}\left[y_{ij}=\mathrm{State\ 2}\right],
    \label{eq:2}
\end{equation}
\begin{equation}
    \mathrm{Difference\ Sum} =\sum_{i=1}^{n}\left[y_{ij}=\mathrm{State\ 3}\right].
    \label{eq:3}
\end{equation}
We now have all the information we need to create \techname.

\techname\ displays each attribute as a grid-like panel (\autoref{fig:wvc}).
Every panel has $n$ adjacent columns, each representing beliefs about an agent. Each column $j$ is composed of the same four components that make up \techname: Ego View, Similarity View, Difference View, and Deltas. 
Every panel also contains a `Delta Line'. Below the Delta Line, the bottom portion of every panel is the `Summary View' that summarizes the system's synchronization state, and the bottom-most component is the `Ego View', which visualizes an agent's ego value.
% to which the bottom-most component of the Summary View is the `Ego View' which visualizes an agent's ego value. 
% The bottom-most component of the Summary View is the `Ego View', which visualizes an agent's ego value.
% To visualize an agent's ego value, the bottom-most component of the Summary View is the `Ego View'.
% showing each agent's ego value. %and the visual representation for an agent's ego value differs based on the attribute.

The visual representation of the ego value is specific to each attribute. For Communication Network, the ego view of each agent's communication network (i.e., the bandwidths from the agent to all other agents) is represented as a 1D-array of length $n$.
The $k$-th entry in the array represents the communication bandwidth value from agent $j$ to agent $k$;  we encode the bandwidth value with a purple-to-red sequential colormap\cite{harrower2003colorbrewer}.
In contrast, the data types used in the Battery Level Panel and Science Zone Panel are ordinal and boolean, respectively; for these, we use a square mark to represent the Ego View. 
Ordinal values are encoded with an orange sequential colormap, and boolean values are encoded with two shades of teal.

Next, above the Ego View, the `Similarity View' displays the similarities of each agent's worldview through a piling metaphor\cite{bach2015small} where the number of horizontal lines represents the Similarity Sum (\autoref{eq:2}).
This information is also captured as a fraction below the Ego View. 
The piling metaphor fits the need of the visualization as it visually aggregates information--reducing visual noises and allowing more emphasis on the differences.
This visual design decision is also based on what we have learned from the formative study, as researchers are more interested in identifying the differences of agents' worldviews rather than the similarities.

% \vspace{-0.05cm}
The next component of \techname\ is the `Difference View', which is adjacent to the Similarity View.
The Difference View is represented by diagonal hashed lines, and it complements the Similarity View.
From the Difference Sum (\autoref{eq:3}), the height of the Difference View indicates how many agents disagree with an agent's ego value (\textbf{R1}). To see who the contrarians are and their beliefs, we look at the top portion of \techname: the Detail View (\textbf{R1, R3}).
Above the Delta Line, the Detail View has $n$ rows representing the $n$ agents.
For each column $j$, rows corresponding to an agent that disagrees with the column's ego view (i.e., rows corresponding to an entry in State 3 in \textbf{Y}) report  the contrarian agent's belief about agent $j$'s worldview. 
For example, in \autoref{fig:wvc} (BT), Agent 7 disagrees with Agent 1 and 4's ego view of their location.
The Detail View displays a different color compared to the respective Ego View's at the bottom. 
By default, rows corresponding to agents that agree with the Ego View are blank, in line with the diff metaphor.
However, users can also show values that correspond to the Ego View by toggling the Similarity View.

Our decision to focus on representing the agents' view of each attribute as a single matrix is based on the lessons from past designs from the co-design sessions.
The past designs layered various visual encoding for each attribute, and users found it difficult to make sense of the layered result and to compare the differences between the agent's worldviews.
In addition, we found that operators would focus on a particular attribute depending on the context of the problem.

\subsection{User Interactions}
\label{sec:interactions}
The prototype features a rich set of user interactions:

\textbf{Highlighting.} To support switching analytical perspectives (TC, AC) when making sense of an agent's schedule, users can highlight a single task or highlight an agent's chain of tasks (\textbf{R2}).
In \autoref{fig:main}, the user highlighted Agent 5's science chain of tasks.
Users can also highlight rows in the Detail View in the DWC as shown in \autoref{fig:wvc}.

\textbf{Interlink Views.}
Visualizations for MRS can be broken down into two aspects: (i) the behavior of a single agent or (ii) the overall behavior of the system\cite{schroeder2001multi}. 
No single tool is capable of providing a complete picture of the system\cite{ndumu1999visualising}.
Hence, we focus on providing the operator with the ability to collate and fuse pieces of information in order to properly diagnose the system by interlinking the views \textbf{(R4)}.
For instance, when the rows in the Communication Network panel are highlighted, the immediate edges of the highlighted agents are pink in the graph, while the opacity of the other edges are lowered. \autoref{fig:root} showcase this interaction.
As another example, the graph, scatterplot, and task abstraction are interlinked.
When an agent is selected in any of  these views, it is simultaneously highlighted in the other views. 

\subsection{System Architecture \& Implementation}
We use a MongoDB database to store each agent's reported worldview.
From the database, we perform three computational tasks before visualizing the system: (i) compute summary statistics; (ii) chain tasks according to their precedence constraints and inter-task dependencies; (iii) compare agents' worldview through the diff function (\autoref{sec:application:dwc}).
Afterward, we visualize the results on a web application.
The front-end is implemented with a combination of HTML5, CSS, JavaScript, and the JavaScript Data-Driven Documents (D3) library\cite{bostock2011d3}.
The back-end runs on a Node.js web server.
\section{System Evaluation}
We are interested in evaluating if MOSAIC Viewer can help operators successfully answer the critical questions listed in \autoref{tab:goals}.
We perform two assessments.
First, we conduct a qualitative study that consists of a training task scenario and three task scenarios in which data were collected.
For each scenario, participants were asked to explore the state of a multi-robot system. 
Then, based on the user study, we present two case studies that demonstrate our system's efficacy and the workflow of how users interacted with the system.

\subsection{User Study}
\label{sec:userstudy}
\textbf{Participants.} We recruited 12 participants (3 female, 9 male), aged reported in bins 18 - 44 years, from the population of multi-robot systems researchers and operators at NASA JPL.
Of the 12 participants, two were part of the formative evaluation. We elected to include these participants in the study as the system's interactions and visual representations changed after formative evaluation, and the study was designed to assess their ability to debug worldviews based on scenarios they did not encounter in the past.
A more comprehensive overview of the participants can be found in \autoref{tab:users}.

\textbf{Conditions and Tasks Design.} To study the usability and outcomes of domain experts using MOSAIC Viewer, we devised three scenarios, each touching upon a common situation operators face. 
One scenario included all agents in sync.
Two scenarios included an ``out of sync'' condition that emerges from (i) one agent isolated from others or (ii)
a bipartition in the agents' communication network, respectively.
For each scenario, we asked operators to assess the system.
If they determined worldviews are out of sync, we asked operators to perform a root cause analysis. 
This involved determining which robots were out of sync and analyzing why they were out of sync \textbf{(Q1)}.
Participants were also asked to assess whether agents accomplished their science tasks \textbf{(Q2)}, navigation tasks \textbf{(Q3)} as well as whether they understood how tasks  were scheduled \textbf{(Q4)}.
Each participant was required to complete the three scenarios, and the study was counterbalanced to mitigate learning effects. 
We chose not to have a baseline interface to compare with MOSAIC Viewer instead of using existing debugging tools for two reasons: (1) reduce confounding effects that may emerge from other aspects of the interfaces and (2) focus the investigation on the qualitative, behavioral aspects participants gained from MOSAIC Viewer.

\begin{figure}[tb]
    \centering
    \includegraphics[width=\columnwidth]{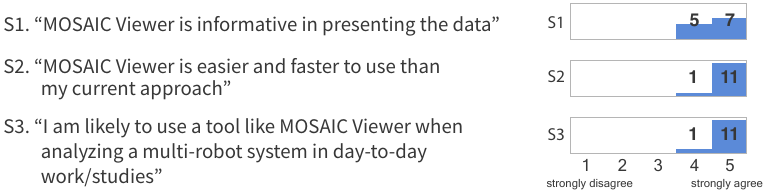}
    \caption{Participant's feedback about MOSAIC Viewer on a 5-point Likert Scale. Median ratings are indicated in gray.}
    \label{fig:responses}
\end{figure}
\textbf{Experimental Setup.} MOSAIC Viewer ran on a mid-2017 MacBook Pro (16 GB, 2.5 GhZ process). The interface was displayed on a 34-inch display (3440 $\times$ 1440 pixels), using the Google Chrome browser. 
Participant input was captured through an external keyboard and mouse.
For the sake of uniformity, pen and paper were provided to participants regardless of the task.

\textbf{Procedure.} Each participant first filled in their background information in a survey form.
They were then trained on the interface until they were comfortable using it. 
Once ready, we provided the three scenarios one by one. 
For each scenario, the participants were asked to answer the four aforementioned questions and write down their answers on the provided sheet of paper. 
At the end of each task, participants responded to a questionnaire about the confidence of their answers.
After finishing all three tasks, participants answered additional survey questions that prompted their overall thoughts about the system and engaged in a semi-structured exit interview.
We used a concurrent think-aloud protocol during the study, and participants were audio- and screen-recorded for the duration of the tasks.
All of the participant's answers for the tasks were saved.

\subsection{Results}
Our evaluation revealed the following findings:
\begin{compactitem}
    \item \textbf{Speed and interactivity streamline higher-level analyses}:
   (1) MOSAIC Viewer helps users locate information faster; (2) The interlinked views enable users to formulate a hypothesis about the root cause of the desynchronized worldviews.
    \item \textbf{Trust for summary displays grew with experience}:
    Users initially lacked trust in the visual representations to which they were not accustomed, but this trust grew quickly once they were able to verify their understanding.
    \item \textbf{The why: The how is not enough on its own}:
    Understanding MRS agents requires not only understanding how agents are interacting with each other and how the tasks are scheduled but also why. One without the other is not complete. 
    \item \textbf{Different sets of assumptions affect data interpretation}:
    Users bring in a different set of assumptions that mismatches from the system's architecture. These mismatches led to incorrect data interpretation.
\end{compactitem}
P6 explains that MOSAIC Viewer meets an unmet need with existing tools, noting ``Every time I build a new capability...There are no tools out of the box that just does it for you. You have to go spend time and build it, so you can properly visualize what our algorithms do.''

\subsubsection{Speed and interactivity streamline analyses}
\textbf{Navigating information.}
All 12 users reported that understanding the Main View’s visual encoding required very low to low-level of effort. P7 attributes the high learnability due to the fact ``someone actually thought about how to represent these data rather than [users] just plotting the data in a given software''.
In particular, participants reported the graph and shapes to be intuitive and useful.

8 out of 12 participants helped explain this finding, sharing that their current workflows required them to open multiple terminal windows per agent, often spreading across multiple monitors to unpack what even a single agent is doing.
This matches the finding from the formative study.
Even when working alone, participants describe terminal logs as time-consuming and inefficient, especially for comparisons. P1 explains this workflow requires visually scanning and remembering the contents--making it mentally taxing. P2 contrasts this multi-screen collaboration experience with MOSAIC Viewer’s compact form, observing that ``[in MOSAIC Viewer] whatever I want to see, the information is there.'' 

Users contrasted MOSAIC Viewer with terminal logs, which require considerable work from users to find individual pieces of information, let alone combine them into a higher-level analysis. P1 comments,
``that isn’t possible with the way I [currently] do it''. Based on the feedback users provided on a 5-point Likert scale, participants describe MOSAIC Viewer as both easier- and faster-to-use than their current approaches ($Md =5$, $IQR =0$,
\begin{sparkline}{3}\sparkrectangle 0.0 1.0
\sparkspike .083 0.00
\sparkspike .283 0.00
\sparkspike .483 0.00
\sparkspike .683 0.09
\sparkspike .883 1.00
\end{sparkline}) (\autoref{fig:responses}). Others found the flexibility helped drive ease-of-use, with P2 observing ``you have more than one way to see [the same] information''.

\textbf{Formulating.}
Previous research explains that debugging a distributed multi-robot system requires users to combine macro- (i.e., societal-level) and micro-level (i.e., agent-level) system state information to form a coherent, unified picture\cite{ndumu1999visualising}.
In our study, users explain that the speed of specific data access enables them to more quickly achieve these higher levels of understanding.
To complete each task, users would use the \techname\ to identify if an agent is out of sync with the other agents. 
To explain a de-synchronization, users would have to combine their domain knowledge and the information provided from  each of the various views to conduct a root cause analysis (\textbf{R4}). 
Users would look at each view to determine if the agent's location, distance, communication bandwidth, or any other variables could explain the desynchronization they determined in \techname. 
In \autoref{sec:case}, we provide a walk-through of how users utilized our system with two case studies.

\begin{figure}[tb]
	\centering
	\captionsetup{farskip=0pt}% <--- no gap at the 
    \includegraphics[width=0.5\linewidth]{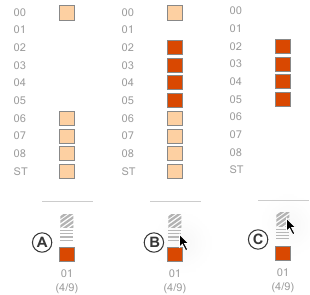}
    \caption{Toggle interaction used in the \techname. (a) show the default setting, only showing the deltas in the Detail View. (b) shows when P4 toggles the Similarity View and the agents that agree with Agent 1’s ego view of its battery level appears in the Detail View. (c) shows when P4 toggles the Difference View and makes the deltas disappear.}
	\label{fig:trust-interactions}
\end{figure}

\subsubsection{Trust for summary displays grew with experience}
Independent of which task was presented first, users first interacted with the system in a way that indicated they were verifying the low-level details in which the summary displays were based upon.

The \techname\ Summary View provides a good example of this observed user behavior. 
The default setting of \techname\ shows only the differences in the Detail View, but users can toggle different parts of the Summary View to see different low-level details (\autoref{fig:trust-interactions}).
Another example is the Task Abstraction view. 
The Task Abstraction view utilizes filled-in shapes to represent the three step process of the science and navigation chain of tasks; however the shapes do not indicate when or who has executed the respective task.
To verify the abstraction, users can individually click on a shape and see the particular task being highlighted in the timeline visualization or automatically highlight the entire chain of tasks as shown in \autoref{fig:wvc}.
\begin{figure}[tb]
	\centering
	\captionsetup{farskip=0pt}% <--- no gap at the 
    \includegraphics[width=\linewidth]{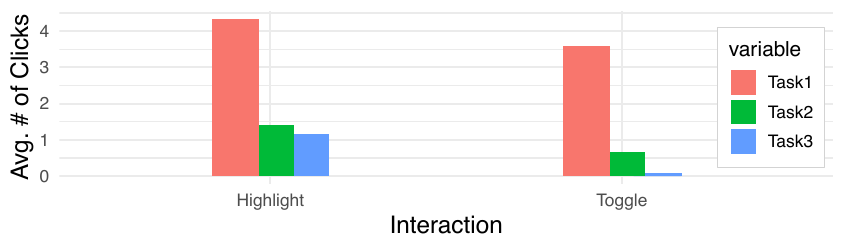}
    \caption{Summary of participants' average number of clicks for two interactions. Participants' average number of clicks decreases over the study. The average for the toggle interaction in the third task is 0.}
	\label{fig:barchart}
	\vspace{-0.1cm}
\end{figure}

\autoref{fig:barchart} shows the average number of clicks users used to either highlight and toggle throughout the user study's three tasks.
In the first task, 8 of 12 users used the toggle interaction to validate their understanding of the \techname’s visual encoding (\autoref{fig:trust-interactions}).
As P4 explains: ``The reason why I've been [toggling] is because I'm not sure right now whether I'm seeing the similarity or differences (in the Detail View). The way I can check that is by looking at the color of [Agent 1's ego view of battery level], and toggling the [diagonal lines] to confirm what I'm seeing.''
By the second task, only 3 users continued to use the toggle interaction to validate their understanding, and 0 users used the toggle interaction by the third task. 
As P6 explains, ``At first, I [toggled] because I wanted to see the whole picture. But once I got used to the system, I don’t need to verify. I know from here it's going to show the same thing so once you get used to it, it's not necessary.'' 

A similar pattern of use was observed when users were asked whether agents have completed their science objectives. During the first task, 7 out of 12 users used the chain of task highlighting interaction to confirm whether an agent's science task abstraction was accurately reflected in the timeline visualization. 
By the third task, only 3 used the interaction, while the rest used the Task Abstraction exclusively. P8 commented: ``It took me a while to really (pause) trust that the little [shapes] were fully representative (of the timeline visualization). I just wanted to double-check that's the right answer.''

\subsubsection{What and how are not enough: expert users need to know why}
While participants shared how MOSAIC Viewer successfully explains what the agents are doing and provide insight to some anomalous behaviors, expert operators wanted even more detail than the system provided. P6 provides a representative quote:
\begin{quote}
    \vspace{-0.1cm}
    Usually, people using [the system] not only want to see how it’s scheduled. I think [answering] how [the tasks] are scheduled, the visualizations are doing that beautifully. But there’s some underlying optimization going on where the objective function is trying to maximize something...As a human operating a system, one of the things we always want to do is verification. So my computer is telling me this [solution] is the best. Is it really the best? Can I check the \textbf{sanity} of the solution?
    \vspace{-0.1cm}
\end{quote}

Similarly, 10 participants indicated they would like to know more about the reasoning behind the scheduling optimization with comments such as, ``I feel like 7 and [base station] should be doing better than that and I wonder why they're not'', ``I want to know why [Agent 5’s science task] was transferred to 4. Why not to 0 or 2?'', ``Why are there blank spaces in the timeline?’’.
During the in-depth interviews, participants commented that they would like to see explanations behind the scheduler’s decisions.
For example, when an agent's task is delegated to another agent, P6 and P10 explain they would want to see the low-level details, such as real-time CPU load, of the two agents in order to understand the scheduling optimization. 

\subsubsection{Different sets of assumptions affect sensemaking}
Despite an unbounded training period, participants who are not familiar with MOSAIC’s specific logistics displayed behavior where they interpreted data differently. We note two specific cases.

P7 and P9, two non-core MOSAIC participants, incorrectly answered ``no''  for all three tasks when asked if the agents accomplished their navigation task (\textbf{Q3}). 
In their think-aloud process, both mentioned the squares that represent the base station's navigation chain of task in the Task Abstraction are not filled in, indicating that the base station failed to accomplish its navigation tasks.
However, as explained in \autoref{sec:mosaic}, the base station is a special agent that does not need to accomplish the mandatory navigation task as its purpose is to help other agents with its fast computation power.

A similar pattern of mental model mismatch occurred with P4.
During the first task, P4 provided an incorrect answer to the question related to the agents' science tasks (\textbf{Q2}).
In their think-aloud process, P4 observed how Agent 3 accomplished its science chain of task with the chain of task highlighting and stated the answer to Q2 is ``yes''.
However, according to the Task Abstraction in \autoref{fig:main}, Agent 5 did not accomplish its science task.
From the exit interview, P4 elaborated the reasoning behind their answer is based on the idea of how ``the agents are working together in tandem'' where if one agent accomplished its science chain of tasks, other agents did as well.
\begin{figure*}[tb]
    \centering
    \includegraphics[width=\textwidth]{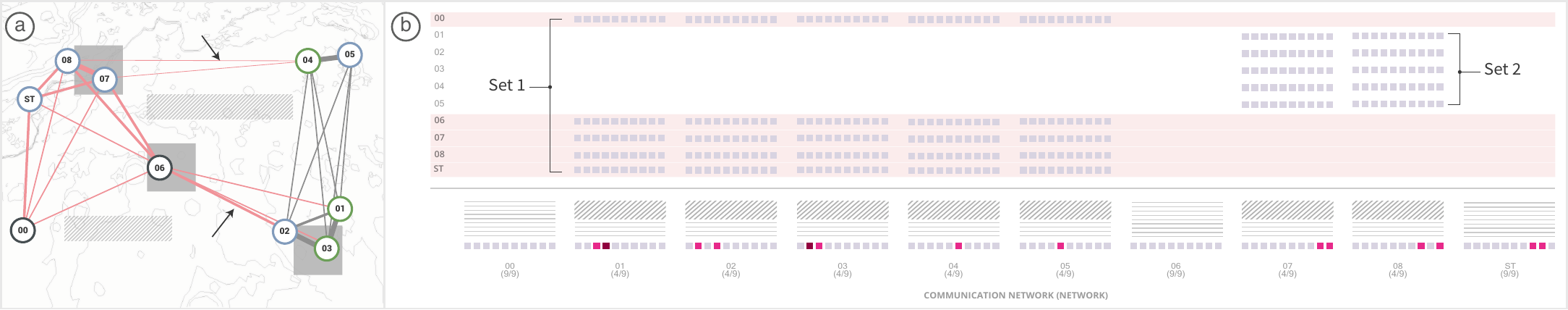}
    \caption{(a) Graph and (b) Communication Network Panel. In (b), operator highlights row 0, 6, 7, 8, ST and identifies a visual pattern of how Set 1 is the complement set of Set 2. They hypothesize there's a network bipartition. The immediate edges of these selected agents are highlighted in (a) and the same group of agents are separated. The weak communication links (indicated by black arrows) further support this hypothesis.}
    \label{fig:root}
    \vspace{-0.5em}
\end{figure*}

\subsection{Case Studies}
\label{sec:case}
\textbf{Evaluate System Performance.} Upon launching the system, the operator sees the Main View and DWC side-by-side. The Summary View (\autoref{fig:main}a) does not display a synchronization warning and DWC also does not show any deltas. The operator concludes the agents’ worldviews are in sync and moves on to assess the system's performance (i.e., how many tasks are being performed, and by whom) as part of their next goal (\autoref{tab:goals}).  

The operator looks at the Task Abstraction (\autoref{fig:main}d) to obtain a high-level overview of the system’s performance. The  filled-in squares in the ``NAV'' column indicate agents have accomplished their mandatory navigation task. The ``SCI'' column shows Agent 3, with its three filled-in triangles, is the only agent that fully accomplished its science chain of tasks.
However, the Science Objectives (\autoref{fig:main}a) and the graph (\autoref{fig:main}c) show there are a total of four eligible agents (Agents 1, 3, 5, and 6) that can accomplish science tasks. 

The operator engages with the timeline (\autoref{fig:main}e) and scatterplot (\autoref{fig:main}b) to seek for the low-level details that explain why the other three agents failed to accomplish their science tasks. 
Highlighting Agent 5’s partial science chain of tasks (\autoref{fig:main}e colored in pink), the operator sees Agent 5 relocated the second science task to Agent 4. 
From the graph, the shortest path to send the scientific data from Agent 4 to the base station is through Agent 8 or Agent 1. However, Agent 4 has a weak communication link to both agents and is unable to relay the data within the time gap from $t=23$ to $t=27$.
Next, the scatterplot shows that Agent 1's and Agent 6's battery level is 50\%, indicating the computation time needed for their tasks will take longer compared to other agents (e.g., Agent 1 takes twice as long as Agent 0 to accomplish its second navigation task).
Evidently, Agent 1 and Agent 6 (colored in orange and turquoise, respectively) did not accomplish any science tasks as both needed more time to accomplish their mandatory navigation tasks.

\textbf{Plan for Corrective Action.}
With the same ``out of sync'' scenario from the user study, this case study reflects how operators can plan for corrective action (i.e., determine who is out of sync and reason why the desynchronization occurred) with MOSAIC Viewer. 

Launching the system, the operator sees the Main View and DWC side-by-side. The Main View signals a desynchronization warning in the Summary Overview, and DWC (\autoref{fig:wvc}) displays deltas across all three panels. The deltas specifically in the Communication Network Panel (\autoref{fig:root}b) show two distinct sets. Highlighting rows 0, 6, 7, 8, ST in the Communication Network panel emphasizes the visualization’s white space, revealing a bipartition in the communication network as Agents 0, 6, 7, 8, ST are out of sync with Agents 1-5, and conversely Agents 1-5 are out of sync with Agent 7 and 8. The deltas in the other two panels have the same visual pattern and the same set of agents (Agents 0, 6, 7, 8, and ST) are out of sync.

The pattern revealed by the DWC is also displayed in the graph in the Main View (\autoref{fig:root}a): the same grouping of agents in \techname\ is reflected within the two distinct clusters of the graph. The weak communication bandwidth between the two clusters (indicated by black arrows) further supports the hypothesis of a bipartite graph. For instance, the bandwidth value between Agent 4 and 7 is  1, a weak value where agents would not be able to send data across the network and update each others’ worldviews, causing the bipartition.

\section{Discussion}
In this section, we reflect upon the findings of this work, and consider both its extension, and its limitations.

\textbf{Limitations in Visual Scalability}.
While the scale of ten agents in our work aligns with current and proposed space exploration concepts\cite{board2012vision,Mannucci10, Ref:OPAL, Ref:CYGNSS} within the next 10 years, the current system design might struggle to achieve the same levels of legibility and low cognitive load with swarm robotics\cite{navarro2012introduction} mission concepts that include hundreds of agents.
Edge bundling\cite{gansner2011multilevel} and layout algorithms \cite{kieffer2015hola} are promising directions to improve dense graph legibility and transparency\cite{dang2010stacking}, and jitter\cite{trutschl2003intelligently} 
can improve dense scatterplot legibility. However, critically, in swarm robotics applications, agents are rarely endowed with a global worldview---rather, each agent typically relies on much simpler scheduling schemes and reacts based on its own state and immediate neighbors'. Accordingly, we expect that further research will be required to capture the qualitatively different nature of the underlying problem of debugging autonomous swarms' behaviors.

\textbf{Foreground scheduler logistics to achieve generalizability}.
In our findings, we observed differences between how core MOSAIC and non-core researchers interpreted the same visual marks on the display. 
In the example reported in Sect 6.2.4, the Task Abstraction shows unfilled squares for the base station.
Some non-core team members interpreted that not all robots have completed their navigation tasks when seeing unfilled squares for the base station. However, core team members saw the same marks, and, knowing that the base station does not need to complete the navigation task, made the correct interpretation that all mandatory navigation tasks were completed. Reflecting on the systems' visual encoding, we realize this confusion could have been avoided by adding an additional glyph that distinguishes the base station from other agents.

This split in our test population reveals an interesting artifact of our design process. Though we spent nearly a year on iteratively designing, we never observed this problem in our user testing until we included users outside of the core MOSAIC team. We reflect that while both groups contain full-time MRS robotics researchers, the core group is more deeply steeped in the architecture and algorithms that drive the MOSAIC scheduler. 
To design a more general MRS application that would work across any number of scheduling algorithms, this would require to dive deep into the mechanics of scheduling algorithms. 
A general system would need to visualize not only the various states of the worldviews, but also many of the underlying mechanisms through which the scheduling system operates. With this approach, designers could be more likely to create a system that functions independently of the background of its users.

\textbf{Move Beyond the Debugging Use Case.} 
This work focuses more explicitly on explaining \textit{what} the autonomy has decided. 
Given this success of this work, participants also expressed some desire to understand, as P7 states, ``different perspectives of the same code'', that extends to understanding \textit{why} the autonomy made the decisions it did in non-error cases. This points towards supporting the task of scheduling algorithm design, such as adding a ``what-if'' mode\cite{wexler2019if}, to help roboticists gain insight into tasks that include how different algorithm hyperparameters affect the search of a very large space. This creates an opportunity to work in algorithm visualization\cite{grissom2003algorithm, shaffer2010algorithm}, as well as large decisions trees visualization \cite{szekely2008scheduling, taylor2002vista}

\textbf{Extend to other Multi-Robot Systems.}
While this system was designed to support users of the MOSAIC scheduler, it also represents an interesting challenge to extend the application to other MRS systems that use different underlying algorithms and for different contexts of use. In one exit interview, for example, P5 observed that MOSAIC Viewer might also support agent navigation discrepancies in support of DARPA's Subterranean Challenge (SubT)\cite{Ref:Darpa}. P5 reflected on ways in which the contexts share many commonalities that could help the systems achieve more generalizability, noting ``What you’re visualizing is at the core of these multi-robot problems. They are connected by some network infrastructure, they’re sharing information in order to come to a decision about what to do. So we want to visualize what were the states at a specific time, what were they doing, what were they planning on doing, and how they were connected with each other.'' But they also noted other system differences that might not make the contexts topologically equivalent, noting ``[SubT’s] task network is a bit larger [than MOSAIC’s] and not as straightforward as 1, 2, 3. You roll back and there’s loops. So it’s harder to just lay it out sequentially like [MOSAIC’s]''.
\section{Conclusion}
We present MOSAIC Viewer, a visual analytics system that helps users make sense of robots' autonomous scheduling decisions and pinpoint the cause of desynchronized worldviews in a multi-robot system.
To compare worldviews, we draw inspiration from the \textit{diff} algorithm to visually emphasize the differences, while aggregating the similarities.
This approach allows users to quickly detect the differences and similarities of all the robots' worldviews.
The interlinked views help users not only collate and fuse pieces of information from each view in order to conduct a root cause analysis of the desynchronized worldviews, but also understand the behavior of the system at a societal and individual level.
Our qualitative user study with domain experts at the NASA JPL characterizes and elaborates the usefulness and effectiveness of MOSAIC Viewer.
\section{Acknowledgement}
The development of MOSAIC Viewer was enabled by the JPL/Caltech/ArtCenter data visualization program. We would like to thank Santiago Lombeyda, Hillary Mushkin, Maggie Hendrie, Alessandra Fleck, and Sarah Strickler for their feedback and contribution on the earlier prototypes of MOSAIC Viewer. Also, special thanks to the MOSAIC team!
This research is sponsored in part by the U.S. National Science Foundation through grant IIS-1741536.
The research was carried out at the Jet Propulsion Laboratory, California Institute of Technology, under a contract with the National Aeronautics and Space Administration (80NM0018D0004).

%\acknowledgments{}

\clearpage
% BALANCE COLUMNS
\balance{}

\bibliographystyle{abbrv-doi}
\bibliography{00_main}

\begin{thebibliography}{10}

\bibitem{altheide2008emergent}
D.~Altheide, M.~Coyle, K.~DeVriese, and C.~Schneider.
\newblock Emergent qualitative document analysis.
\newblock {\em Handbook of Emergent Methods}, pp. 127--151, 2008.

\bibitem{annable2013nubugger}
B.~Annable, D.~Budden, and A.~Mendes.
\newblock Nubugger: A visual real-time robot debugging system.
\newblock In {\em Proc. RoboCup}, pp. 544--551. Springer, 2013.

\bibitem{arai2002advances}
T.~Arai, E.~Pagello, L.~E. Parker, et~al.
\newblock Advances in multi-robot systems.
\newblock {\em IEEE Transactions on Robotics and Automation}, 18(5):655--661,
  2002.

\bibitem{argrow2005uav}
B.~Argrow, D.~Lawrence, and E.~Rasmussen.
\newblock {UAV} systems for sensor dispersal, telemetry, and visualization in
  hazardous environments.
\newblock In {\em Proc. AIAA Aerospace Sciences Meeting and Exhibit}, p. 1237,
  2005.

\bibitem{bach2015small}
B.~Bach, N.~Henry-Riche, T.~Dwyer, T.~Madhyastha, J.-D. Fekete, and
  T.~Grabowski.
\newblock {Small MultiPiles}: Piling time to explore temporal patterns in
  dynamic networks.
\newblock {\em Computer Graphics Forum}, 34(3):31--40, 2015.

\bibitem{barnes1988developing}
D.~J. Barnes, M.~T. Russell, and M.~C. Wheadon.
\newblock Developing and adapting {UNIX} tools for workstations.
\newblock In {\em EUUG Conference Proceedings}, pp. 321--333, 1988.

\bibitem{baxter2007multi}
J.~L. Baxter, E.~Burke, J.~M. Garibaldi, and M.~Norman.
\newblock Multi-robot search and rescue: A potential field based approach.
\newblock In {\em Autonomous Robots and Agents}, pp. 9--16. Springer, 2007.

\bibitem{board2012vision}
S.~S. Board, N.~R. Council, et~al.
\newblock {\em Vision and voyages for planetary science in the decade
  2013-2022}.
\newblock National Academies Press, 2012.

\bibitem{bostock2011d3}
M.~Bostock, V.~Ogievetsky, and J.~Heer.
\newblock D$^3$ data-driven documents.
\newblock {\em IEEE Transactions on Visualization and Computer Graphics},
  17(12):2301--2309, 2011.

\bibitem{burmeister1997application}
B.~Burmeister, A.~Haddadi, and G.~Matylis.
\newblock Application of multi-agent systems in traffic and transportation.
\newblock {\em IEE Proceedings - Software Engineering}, 144(1):51--60, 1997.

\bibitem{cook2005illuminating}
K.~A. Cook and J.~J. Thomas.
\newblock Illuminating the path: The research and development agenda for visual
  analytics.
\newblock Technical report, Pacific Northwest National Lab., Richland, WA
  (United States), 2005.

\bibitem{cummings2005managing}
M.~Cummings and P.~Mitchell.
\newblock Managing multiple {UAV}s through a timeline display.
\newblock In {\em Proc. AIAA Info Tech}, p. 7060. 2005.

\bibitem{dang2010stacking}
T.~N. Dang, L.~Wilkinson, and A.~Anand.
\newblock Stacking graphic elements to avoid over-plotting.
\newblock {\em IEEE Transactions on Visualization and Computer Graphics},
  16(6):1044--1052, 2010.

\bibitem{de2007distributed}
M.~De~Rosa, J.~Campbell, P.~Pillai, S.~Goldstein, P.~Lee, and T.~Mowry.
\newblock Distributed watchpoints: Debugging large multi-robot systems.
\newblock In {\em Proc. ICRA}, pp. 3723--3729. IEEE, 2007.

\bibitem{Ref:Darpa}
{Defense Advanced Research Projects Agency}.
\newblock {DARPA Subterranean (SubT) Challenge}.
\newblock URL:
  \url{https://www.darpa.mil/program/darpa-subterranean-challenge}.

\bibitem{elmqvist2015patterns}
N.~Elmqvist and J.~S. Yi.
\newblock Patterns for visualization evaluation.
\newblock {\em Information Visualization}, 14(3):250--269, 2015.

\bibitem{emslie1975theory}
A.~Emslie, R.~Lagace, and P.~Strong.
\newblock Theory of the propagation of uhf radio waves in coal mine tunnels.
\newblock {\em IEEE Transactions on Antennas and Propagation}, 23(2):192--205,
  1975.

\bibitem{entin1999adaptive}
E.~E. Entin and D.~Serfaty.
\newblock Adaptive team coordination.
\newblock {\em Human factors}, 41(2):312--325, 1999.

\bibitem{Ref:CYGNSS}
{eoPortal}.
\newblock {CYGNSS (Cyclone Global Navigation Satellite System)}.
\newblock URL:
  \url{https://directory.eoportal.org/web/eoportal/satellite-missions/c-missions/cygnss}.

\bibitem{Ref:OPAL}
{eoPortal}.
\newblock {OPAL (Orbiting Picosatellite Automatic Launcher)}.
\newblock URL:
  \url{https://directory.eoportal.org/web/eoportal/satellite-missions/o/opal}.

\bibitem{fails2006visual}
J.~A. Fails, A.~Karlson, L.~Shahamat, and B.~Shneiderman.
\newblock A visual interface for multivariate temporal data: Finding patterns
  of events across multiple histories.
\newblock In {\em Proc. VAST}, pp. 167--174. IEEE, 2006.

\bibitem{figueiredo2006multi}
J.~Figueiredo, N.~Lau, and A.~Pereira.
\newblock Multi-agent debugging and monitoring framework.
\newblock {\em IFAC Proceedings Volumes}, 39(20):114--120, 2006.

\bibitem{gancet2010user}
J.~Gancet, E.~Motard, A.~Naghsh, C.~Roast, M.~M. Arancon, and L.~Marques.
\newblock User interfaces for human robot interactions with a swarm of robots
  in support to firefighters.
\newblock In {\em Proc. ICRA}, pp. 2846--2851. IEEE, 2010.

\bibitem{gansner2011multilevel}
E.~R. Gansner, Y.~Hu, S.~North, and C.~Scheidegger.
\newblock Multilevel agglomerative edge bundling for visualizing large graphs.
\newblock In {\em Proc. PacificVis}, pp. 187--194. IEEE, 2011.

\bibitem{garcia1984debugging}
H.~Garcia-Molina, F.~Germano, and W.~H. Kohler.
\newblock Debugging a distributed computing system.
\newblock {\em IEEE Transactions on Software Engineering}, (2):210--219, 1984.

\bibitem{gleicher2011visual}
M.~Gleicher, D.~Albers, R.~Walker, I.~Jusufi, C.~D. Hansen, and J.~C. Roberts.
\newblock Visual comparison for information visualization.
\newblock {\em Information Visualization}, 10(4):289--309, 2011.

\bibitem{gove2014seem}
R.~Gove, J.~Saxe, S.~Gold, A.~Long, and G.~Bergamo.
\newblock {SEEM}: a scalable visualization for comparing multiple large sets of
  attributes for malware analysis.
\newblock In {\em Proc. VizSec}, pp. 72--79. IEEE, 2014.

\bibitem{grissom2003algorithm}
S.~Grissom, M.~F. McNally, and T.~Naps.
\newblock Algorithm visualization in {CS} education: Comparing levels of
  student engagement.
\newblock In {\em Proc. SoftVis}, pp. 87--94. ACM, 2003.

\bibitem{halloran2006unfolding}
J.~Halloran, E.~Hornecker, G.~Fitzpatrick, M.~Weal, D.~Millard, D.~Michaelides,
  D.~Cruickshank, and D.~De~Roure.
\newblock Unfolding understandings: Co-designing ubicomp in situ, over time.
\newblock In {\em Proc. DIS}, pp. 109--118. ACM, 2006.

\bibitem{halpern1990knowledge}
J.~Y. Halpern and Y.~Moses.
\newblock Knowledge and common knowledge in a distributed environment.
\newblock {\em Journal of the ACM}, 37(3):549--587, 1990.

\bibitem{harrower2003colorbrewer}
M.~Harrower and C.~A. Brewer.
\newblock {ColorBrewer.org}: An online tool for selecting colour schemes for
  maps.
\newblock {\em Cartographic Journal}, 40(1):27--37, 2003.

\bibitem{holland2000emergence}
J.~H. Holland.
\newblock {\em Emergence: From chaos to order}.
\newblock OUP Oxford, 2000.

\bibitem{holtzblatt1997contextual}
K.~Holtzblatt and H.~Beyer.
\newblock {\em Contextual design: Defining customer-centered systems}.
\newblock Elsevier, 1997.

\bibitem{humphrey2006visualization}
C.~M. Humphrey, S.~M. Gordon, and J.~A. Adams.
\newblock Visualization of multiple robots during team activities.
\newblock In {\em Proc. HFES}, vol.~50, pp. 651--655. SAGE Publications, 2006.

\bibitem{hunt1976algorithm}
J.~W. Hunt and M.~D. MacIlroy.
\newblock {\em An algorithm for differential file comparison}.
\newblock Bell Laboratories Murray Hill, 1976.

\bibitem{jin2008vizscript}
J.~Jin, R.~Sanchez, R.~T. Maheswaran, and P.~Szekely.
\newblock Vizscript: On the creation of efficient visualizations for
  understanding complex multi-agent systems.
\newblock In {\em Proc. IUI}, pp. 40--49. ACM, 2008.

\bibitem{jo2014livegantt}
J.~Jo, J.~Huh, J.~Park, B.~Kim, and J.~Seo.
\newblock {LiveGantt}: Interactively visualizing a large manufacturing
  schedule.
\newblock {\em IEEE Transactions on Visualization and Computer Graphics},
  20(12):2329--2338, 2014.

\bibitem{kardas2012design}
G.~Kardas, M.~Challenger, S.~Yildirim, and A.~Yamuc.
\newblock Design and implementation of a multiagent stock trading system.
\newblock {\em Software: Practice and Experience}, 42(10):1247--1273, 2012.

\bibitem{karras2017puffer}
J.~T. Karras, C.~L. Fuller, K.~C. Carpenter, A.~Buscicchio, D.~McKeeby, C.~J.
  Norman, C.~E. Parcheta, I.~Davydychev, and R.~S. Fearing.
\newblock Pop-up mars rover with textile-enhanced rigid-flex {PCB} body.
\newblock In {\em Robotics and Automation (ICRA), 2017 IEEE International
  Conference on}, pp. 5459--5466. IEEE, 2017.

\bibitem{kasper2019sun}
J.~Kasper, J.~Lazio, A.~Romero-Wolf, J.~Lux, and T.~Neilsen.
\newblock The sun radio interferometer space experiment (sunrise) mission
  concept.
\newblock In {\em 2019 IEEE Aerospace Conference}, pp. 1--11. IEEE, 2019.

\bibitem{kieffer2015hola}
S.~Kieffer, T.~Dwyer, K.~Marriott, and M.~Wybrow.
\newblock Hola: Human-like orthogonal network layout.
\newblock {\em IEEE Transactions on Visualization and Computer Graphics},
  22(1):349--358, 2015.

\bibitem{kunsei2018improved}
H.~Kunsei, K.~S. Bialkowski, M.~S. Alam, and A.~M. Abbosh.
\newblock Improved communications in underground mines using reconfigurable
  antennas.
\newblock {\em IEEE Transactions on Antennas and Propagation},
  66(12):7505--7510, 2018.

\bibitem{liu2016visualizing}
S.~Liu, D.~Maljovec, B.~Wang, P.-T. Bremer, and V.~Pascucci.
\newblock Visualizing high-dimensional data: Advances in the past decade.
\newblock {\em IEEE Transactions on Visualization and Computer Graphics},
  23(3):1249--1268, 2016.

\bibitem{luo2002multi}
Y.~Luo, K.~Liu, and D.~N. Davis.
\newblock A multi-agent decision support system for stock trading.
\newblock {\em IEEE Network}, 16(1):20--27, 2002.

\bibitem{Mannucci10}
A.~J. Mannucci, J.~Dickson, C.~Duncan, and K.~Hurst.
\newblock {GNSS} geospace constellation ({GGC}): A {CubeSat} space weather
  mission concept.
\newblock Technical report, Jet Propulsion Laboratory, California Institute of
  Technology, 2010.

\bibitem{michel2018multi}
F.~Michel, J.~Ferber, and A.~Drogoul.
\newblock Multi-agent systems and simulation: A survey from the agent
  commu-nity’s perspective.
\newblock In {\em Multi-Agent Systems}, pp. 17--66. CRC Press, 2018.

\bibitem{nagatani2011multirobot}
K.~Nagatani, Y.~Okada, N.~Tokunaga, S.~Kiribayashi, K.~Yoshida, K.~Ohno,
  E.~Takeuchi, S.~Tadokoro, H.~Akiyama, I.~Noda, et~al.
\newblock Multirobot exploration for search and rescue missions: A report on
  map building in {RoboCupRescue} 2009.
\newblock {\em Journal of Field Robotics}, 28(3):373--387, 2011.

\bibitem{navarro2012introduction}
I.~Navarro and F.~Mat{\'\i}a.
\newblock An introduction to swarm robotics.
\newblock {\em Isrn robotics}, 2013.

\bibitem{ndumu1999visualising}
D.~T. Ndumu, H.~S. Nwana, L.~C. Lee, and J.~C. Collis.
\newblock Visualising and debugging distributed multi-agent systems.
\newblock In {\em Proc. AAMAS}, pp. 326--333, 1999.

\bibitem{niederer2017taco}
C.~Niederer, H.~Stitz, R.~Hourieh, F.~Grassinger, W.~Aigner, and M.~Streit.
\newblock {TACO}: visualizing changes in tables over time.
\newblock {\em IEEE Transactions on Visualization and Computer Graphics},
  24(1):677--686, 2017.

\bibitem{osawa1996robocup}
E.~Osawa, H.~Kitano, M.~Asada, Y.~Kuniyoshi, and I.~Noda.
\newblock {RoboCup}: the robot world cup initiative.
\newblock In {\em Proc. ICMAS}, pp. 9--13, 1996.

\bibitem{parker2007distributed}
L.~E. Parker.
\newblock Distributed intelligence: Overview of the field and its application
  in multi-robot systems.
\newblock In {\em Proc. AAAI Fall Symposium: Regarding the Intelligence in
  Distributed Intelligent Systems}, pp. 1--6, 2007.

\bibitem{pinciroli2012argos}
C.~Pinciroli, V.~Trianni, R.~O’Grady, G.~Pini, A.~Brutschy, M.~Brambilla,
  N.~Mathews, E.~Ferrante, G.~Di~Caro, F.~Ducatelle, et~al.
\newblock Argos: a modular, parallel, multi-engine simulator for multi-robot
  systems.
\newblock {\em Swarm intelligence}, 6(4):271--295, 2012.

\bibitem{preece1994human}
J.~Preece, Y.~Rogers, H.~Sharp, D.~Benyon, S.~Holland, and T.~Carey.
\newblock {\em Human-Computer Interaction}.
\newblock Addison-Wesley Longman Ltd., 1994.

\bibitem{quigley2009ros}
M.~Quigley, K.~Conley, B.~Gerkey, J.~Faust, T.~Foote, J.~Leibs, R.~Wheeler, and
  A.~Y. Ng.
\newblock {ROS}: An open-source robot operating system.
\newblock In {\em Proc. ICRA Workshop on Open Source Software}, vol.~3, p.~5.
  Kobe, Japan, 2009.

\bibitem{rossi2018}
A.~Rahmani, S.~Bandyopadhyay, F.~Rossi, J.-P. de~la Croix, J.~V. Hook, and
  M.~T. Wolf.
\newblock Space vehicle swarm exploration missions: A study of key enabling
  technologies and gaps.
\newblock In {\em Proc. IAC}, 2019.

\bibitem{rettig1994prototyping}
M.~Rettig.
\newblock Prototyping for tiny fingers.
\newblock {\em Communications of the ACM}, 37(4):21--27, 1994.

\bibitem{rossi2018routing}
F.~Rossi, R.~Zhang, Y.~Hindy, and M.~Pavone.
\newblock Routing autonomous vehicles in congested transportation networks:
  Structural properties and coordination algorithms.
\newblock {\em Autonomous Robots}, 42(7):1427--1442, 2018.

\bibitem{ruff2013human}
H.~A. Ruff and G.~L. Calhoun.
\newblock Human supervision of multiple autonomous vehicles.
\newblock Technical report, Air Force Research Lab Wright-Patterson AFB OH
  Human Effectiveness Directorate, 2013.

\bibitem{sadana2014onset}
R.~Sadana, T.~Major, A.~Dove, and J.~Stasko.
\newblock Onset: A visualization technique for large-scale binary set data.
\newblock {\em IEEE transactions on visualization and computer graphics},
  20(12):1993--2002, 2014.

\bibitem{schroeder2001multi}
M.~Schroeder and P.~Noy.
\newblock Multi-agent visualisation based on multivariate data.
\newblock In {\em Proc. AAMAS}, pp. 85--91, 2001.

\bibitem{seah2005multi}
C.~Seah, M.~Sierhuis, and W.~J. Clancey.
\newblock Multi-agent modeling and simulation approach for design and analysis
  of mer mission operations.
\newblock 2005.

\bibitem{sedlmair2012design}
M.~Sedlmair, M.~Meyer, and T.~Munzner.
\newblock Design study methodology: Reflections from the trenches and the
  stacks.
\newblock {\em IEEE Transactions on Visualization and Computer Graphics},
  18(12):2431--2440, 2012.

\bibitem{shaffer2010algorithm}
C.~A. Shaffer, M.~L. Cooper, A.~J.~D. Alon, M.~Akbar, M.~Stewart, S.~Ponce, and
  S.~H. Edwards.
\newblock Algorithm visualization: The state of the field.
\newblock {\em ACM Transactions on Computing Education}, 10(3):1--22, 2010.

\bibitem{simmons2000coordination}
R.~Simmons, D.~Apfelbaum, W.~Burgard, D.~Fox, M.~Moors, S.~Thrun, and
  H.~Younes.
\newblock Coordination for multi-robot exploration and mapping.
\newblock In {\em Proc. AAAI/IAAI}, pp. 852--858, 2000.

\bibitem{song2012diffmatrix}
H.~Song, B.~Lee, B.~H. Kim, and J.~Seo.
\newblock {DiffMatrix}: Matrix-based interactive visualization for comparing
  temporal trends.
\newblock In {\em Proc. EuroVis (Short Papers)}, 2012.

\bibitem{szekely2008scheduling}
P.~Szekely, R.~Maheswaran, C.~M. Rogers, and R.~Sanchez.
\newblock Scheduling the activities of distributed teams.
\newblock 2008.

\bibitem{szekely2001interfaces}
P.~Szekely, C.~M. Rogers, and M.~Frank.
\newblock Interfaces for understanding multi-agent behavior.
\newblock In {\em Proc. IUI}, pp. 161--166. ACM, 2001.

\bibitem{tanoto2006mpeg}
A.~Tanoto, J.~L. Du, T.~Kaulmann, and U.~Witkowski.
\newblock {MPEG}-4-based interactive visualization as an analysis tool for
  experiments in robotics.
\newblock In {\em Proc. MSV}, pp. 186--192, 2006.

\bibitem{taylor2002vista}
G.~Taylor, R.~M. Jones, M.~Goldstein, R.~Frederiksen, and R.~E. Wray.
\newblock {VISTA}: A generic toolkit for visualizing agent behavior.
\newblock In {\em Proc. CGF}, pp. 29--40, 2002.

\bibitem{trutschl2003intelligently}
M.~Trutschl, G.~Grinstein, and U.~Cvek.
\newblock Intelligently resolving point occlusion.
\newblock In {\em Proc. InfoVis}, pp. 131--136. IEEE, 2003.

\bibitem{tullis1984predicting}
T.~S. Tullis.
\newblock {\em Predicting the usability of alphanumeric displays}.
\newblock PhD thesis, 1984.

\bibitem{van1995debugging}
M.~H. Van~Liedekerke and N.~M. Avouris.
\newblock Debugging multi-agent systems.
\newblock {\em Information and Software Technology}, 37(2):103--112, 1995.

\bibitem{vander2019autonomous}
J.~Vander~Hook, W.~Seto, V.~Nguyen, Z.~Hasnain, L.~Gallagher, T.~Halpin-Chan,
  V.~Varahamurthy, and M.~Angulo.
\newblock Autonomous swarms of high speed maneuvering surface vessels for the
  central test evaluation improvement program.
\newblock In {\em Unmanned Systems Technology XXI}, vol. 11021, p. 110210M.
  International Society for Optics and Photonics, 2019.

\bibitem{vander2019mars}
J.~Vander~Hook, T.~Vaquero, F.~Rossi, M.~Troesch, M.~S. Net, J.~Schoolcraft,
  J.-P. de~la Croix, and S.~Chien.
\newblock Mars on-site shared analytics information and computing.
\newblock In {\em Proc. ICAPS}, vol.~29, pp. 707--715, 2019.

\bibitem{wexler2019if}
J.~Wexler, M.~Pushkarna, T.~Bolukbasi, M.~Wattenberg, F.~Vi{\'e}gas, and
  J.~Wilson.
\newblock The what-if tool: Interactive probing of machine learning models.
\newblock {\em IEEE transactions on visualization and computer graphics},
  26(1):56--65, 2019.

\bibitem{wongsuphasawat2011lifeflow}
K.~Wongsuphasawat, J.~A. Guerra~G{\'o}mez, C.~Plaisant, T.~D. Wang,
  M.~Taieb-Maimon, and B.~Shneiderman.
\newblock {LifeFlow}: visualizing an overview of event sequences.
\newblock In {\em Proc. CHI}, pp. 1747--1756. ACM, 2011.

\bibitem{yamashita2003motion}
A.~Yamashita, T.~Arai, J.~Ota, and H.~Asama.
\newblock Motion planning of multiple mobile robots for cooperative
  manipulation and transportation.
\newblock {\em IEEE Transactions on Robotics and Automation}, 19(2):223--237,
  2003.

\end{thebibliography}
\end{document}